\documentstyle[amsmath,amssymb,11pt,epsfig]{article}
\title{\Large{\bf Proposal for testing lepton univerality in upsilon
decays at a B factory running at $\Upsilon(3S)$}
\thanks{Research under grant FPA2005-01678 and GVACOMP2006-214.}}
\author{Miguel-Angel Sanchis-Lozano\thanks{Email:Miguel.Angel.Sanchis@uv.es}
\vspace{0.1cm}\\
\it Instituto de F\'{\i}sica
Corpuscular (IFIC) and Departamento de F\'{\i}sica Te\'orica \\
\it Centro Mixto Universidad de Valencia-CSIC \\
\it Dr. Moliner 50, E-46100 Burjassot, Valencia (Spain)}

\begin{document}

\date{}
\maketitle

{\small A proposal is presented for detecting new physics at a B factory 
running at the $\Upsilon(3S)$ resonance by testing lepton universality
to the few percent level in the leptonic decays of the $\Upsilon(1S)$ and 
$\Upsilon(2S)$ resonances tagged by the dipion in the chain decay:
$\Upsilon(3S) \to \pi^+\pi^-\Upsilon(1S,2S);\ \Upsilon(1S,2S)\to\ell^+\ell^-,\
\ell=e,\mu,\tau$.}

\vspace{-10.cm}
\begin{flushright}
  IFIC/06-30\\
  FTUV-06-1004\\
  hep-ph/0610046
\end{flushright}
\vspace{9.cm}

\section{Introduction and motivation}
In the quest for new physics beyond the Standard Model (SM)
high luminosity B factories and the LHC will play complementary roles
in the near future.
In particular, the discovery of the long-awaited Higgs boson(s) is one of the
greatest challenges arising with the advent of the so-called LHC
era. In certain extensions of the SM, however, one of the non-standard
Higgs bosons can be quite (even very) light, making
its detection uncertain at the LHC, especially for a Higgs mass
below the $b\bar{b}$ threshold. Conversely, a high luminosity 
B factory would be the ideal place to discover such a light Higgs
boson, or put stringent constraints on its existence in different
models and scenarios. 
Moreover, a super B factory \cite{Akeroyd:2004mj}
can go ahead in this study by performing measurements with an 
unprecedented accuracy even when running at $\Upsilon(4S)$.

The relevance of (radiative) decays of $\Upsilon$ resonances
in the search for a (pseudo)scalar non-standard particle was soon
recognized \cite{Wilczek:1977pj,Haber:1978jt,Ellis:1979jy}. Let us
mention, however, that the Crystal Ball, CLEO, and Argus experimental searches 
\cite{Peck:1984vx,Besson:1985xw,Albrecht:1985qz} have 
yielded negative results so far. (No further confirmation
was found for a narrow state claimed by Crystal Ball with a mass of
around 8.3 GeV.) Basically, in all these searches, a monochromatic
photon was expected but no peak was observed in the photon
spectrum and narrow states were excluded in the analysis.

In this proposal, we consider the possibility of broader
intermediate state, yielding rather non-monochromatic photons 
in the $\Upsilon \to \gamma\ \tau^+\tau^-$ radiative process. Thus, any
photon signal peak would be smeared and probably swallowed up in the
background. Nevertheless, new physics might still show up as a
(slight) breaking of lepton universality (LU), i.e., the branching
fractions (BFs) for the electronic and muonic on the one hand and the
tauonic mode on the other hand, would be (slightly) different
because of a new physics contribution to the latter.

CLEO recently submitted a paper \cite{Besson:2006gj} where the ratio
of the tauonic and muonic BFs is examined
for all three $\Upsilon(1S,2S,3S)$ states. The conclusion was that
LU is respected within the current experimental accuracy ($\sim$10$\%$), 
although the measured tauonic BF turns
out to be systematically larger than the muonic one at the few
percent level.

\begin{table}[hbt]
\setlength{\tabcolsep}{0.4pc}
\caption{Measured leptonic branching fractions
$BF[\Upsilon(nS) \to \ell \ell]$ (in $\%$) and error bars (summed
in quadrature) of $\Upsilon(1S)$, $\Upsilon(2S)$, and
$\Upsilon(3S)$ resonances (obtained from recent CLEO data and
ref.9).\newline} 
\label{FACTORES}

\begin{center}
\begin{tabular}{ccccc}
\hline channel: & $e^+e^-$ & $\mu^+\mu^-$ & $\tau^+\tau^-$ &
$R_{\tau/\ell}(nS)$ \\
\hline $\Upsilon(1S)$ & $2.38 \pm 0.11$ &  & $2.61 \pm 0.13$ &
$0.10 \pm 0.07$\\
\hline $\Upsilon(1S)$ &             & $2.48 \pm 0.06$ & $2.61 \pm
0.13$ &
$0.05 \pm 0.06$\\
\hline $\Upsilon(2S)$ & $1.91 \pm 0.11$ &    & $2.11 \pm 0.15$ &
$0.11 \pm 0.11$ \\
\hline $\Upsilon(2S)$ & & $1.93 \pm 0.17$ & $2.11 \pm 0.15$ &
$0.09 \pm 0.12$ \\
\hline $\Upsilon(3S)$ & $2.18 \pm 0.20$ &    & $2.55 \pm 0.24$ &
$0.17 \pm 0.14$ \\
\hline $\Upsilon(3S)$ & & $2.18 \pm 0.21$ & $2.55 \pm 0.24$ &
$0.17 \pm 0.15$ \\
\hline
\end{tabular}
\end{center}
\end{table}

\begin{figure}
\begin{center}
\includegraphics[width=16pc]{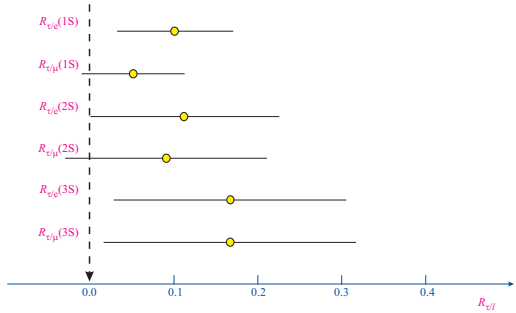}
\end{center}
\caption{\small Plot of $R_{\tau/\ell}$ values from
Table I. According to a hypothesis test
(with LU representing the null hypothesis
predicting $<R_{\tau/\ell}>\ =0$), LU
can be rejected to the $1\%$ level of
significance. A larger precision is required for any claim; however, 
a B factory could provide it.}
\end{figure}

Deviation from LU can be assessed through the ratio defined as 
\cite{Sanchis-Lozano:2005di,Sanchis-Lozano:2003ha}
\begin{equation}
R_{\tau/\ell}(nS)=\frac{BF[\Upsilon(nS) \to \tau
\tau]-BF[\Upsilon(nS) \to \ell \ell]}{BF[\Upsilon(nS) \to
\ell \ell]}=\frac{BF[\Upsilon(nS) \to \tau
\tau]}{BF[\Upsilon(nS) \to \ell \ell]}-1;\ \ \
\ell=e,\mu
\end{equation}

Once these CLEO results are combined with previous BF measurements
(see Table I and Fig.1), one can observe an overall $2.6\sigma$
effect favoring the tauonic decay mode versus both the electron
and muonic ones in all three $\Upsilon(1S,2S,3S)$ resonances,
implying $R_{\tau/\ell}>0$.

As advocated in refs.10-12, the
LU breaking in upsilon decays would hint at new physics
beyond the SM (BSM), pointing out the existence of a light (CP-odd) 
non-standard Higgs boson. Such a hypothetical particle would 
mediate the $b\bar{b}$
annihilation into a tauonic pair subsequent to a dipole magnetic
(M1) transition (see Fig.2), yielding an unobserved (not necessarily
soft) photon according to the process
\begin{equation}
\Upsilon(nS)\ \to\ \gamma\ \eta_b(n'S)(\to \tau^+\tau^-);\ \ n'\leq n
\end{equation}
i.e., we consider both allowed (photon energy
$\lesssim 100$ MeV) and hindered (photon energy $\lesssim 1$ GeV)
transitions between the $\Upsilon$ and $\eta_b$ states.
Factorizing the decay as a two-step process,
\begin{equation}
BF[\Upsilon(nS)\ \to\ \gamma\ \tau^+\tau^-]= BF[\Upsilon(nS)\ \to\
\gamma\ \eta_b] \times BF[\eta_b \to \tau^+\tau^-]
\end{equation}
where the probability of the M1 transition,  
$\Upsilon \to \gamma \eta_b$, can be determined in a potential
quark model calculation \cite{Godfrey:2001eb} or
using an effective theory of QCD \cite{Brambilla:2005zw}.

\begin{figure}
\begin{center}
\includegraphics[width=16pc]{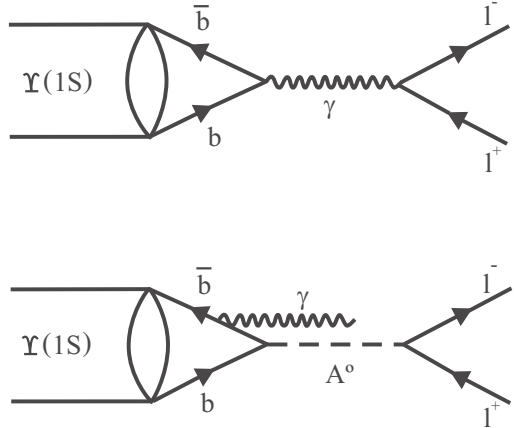}
\end{center}
\caption{\small (a): Conventional electromagnetic annihilation of
the $\Upsilon(1S)$ resonance into a $\ell^+\ell^-$ pair. (b):
Non-standard Higgs-mediated annihilation subsequent to
photon emission either on the continuum or through an intermediate
$b\bar{b}$ bound state. The Higgs-fermion coupling is
proportional to the fermionic mass; hence, only the tauonic decay 
mode should be sensitive to this NP contribution.}
\end{figure}

Leptonic widths and BFs 
as defined in ref.9
contain corrections of all the orders of QED, including radiated photons.
Since the M1 photon in the process (2)
would escape undetected because of the experimental technique
employed, 
the NP contribution would be
unwittingly ascribed to the tauonic decay mode, thereby enhancing
its BF, while the electronic and muonic modes would remain unaltered,
ultimately implying the observation of LU breaking in upsilon decays.
(Notice that higher order SM processes like the $Z^0$ exchange, should give
negligible contributions.)
It is possible, however, to look specifically for the M1 photon
although no clean peak can be expected from continuum
emission or a broad intermediate state.

In summary, assuming universality breaking as a working
hypothesis, the results shown in Table I are compatible with the
following interpretation involving new physics:
\begin{itemize}
\item There is a light CP-odd (or without definite CP) Higgs
particle (denoted as $A^0$) whose mass is at about $10$ GeV.
A significant mixing can occur between the $A^0$ and $\eta_b$ states if the
their masses are close enough. Then, the descriptions based on the
mixing or on an intermediate $\eta_b$ state decaying through
a CP-odd Higgs boson $A^0$ as in Eq.~(2) should be 
equivalent to a leading order approximation.
\item Even for moderate values of $\tan{\beta}$ (defined
as the ratio of the vacuum expectation values of the Higgs
down- and up-doublets \cite{gunion}), the $\eta_b$ resonance would
decay predominantly into a tauonic pair via a Higgs-mediated
annihilation channel (see Fig.2). Hence, a magnetic dipole transition 
from the $\Upsilon$ would yield an
intermediate $\eta_b$ state subsequently decaying into a $\tau^+\tau^-$
pair, with $BF[\eta_b \to \tau^+ \tau^-] \simeq 1$.
\item The BFs shown in Table I are compatible with the probability of 
either allowed or hindered M1 transitions from
$\Upsilon$ resonances into $\eta_b$, whose
estimated BFs \cite{Godfrey:2001eb,Brambilla:2005zw} are both of
the order $10^{-4}-10^{-3}$. 
Thus, since $BF[\Upsilon(nS) \to \ell \ell] \simeq 2\%$, one naturally
obtains
\begin{equation}
R_{\tau/\ell}(nS) \approx\ \frac{BF[\Upsilon(nS)\ \to\ \gamma\
\eta_b]}{BF[\Upsilon(nS) \to \ell \ell]}\ \approx\ 10^{-2}-10^{-1}
\end{equation}
\item In addition,  considering radiative upsilon
decays \cite{Wilczek:1977pj}
into an on-shell $A^0$ particle if kinematically allowed,  
$\Upsilon \to \gamma\ A^0$ occurs, the latter
subsequently decays into a tauonic pair $A^0 \to \tau^+\tau^-$ with
probability near unity for moderate $\tan{\beta}$
\cite{Sanchis-Lozano:2003ha}. 
Setting the reference values
($\tan{\beta}=15$, $v=246$ GeV and 
$M_{\Upsilon}-M_{A^0} = 250$ MeV), one obtains
\begin{equation}
R_{\tau/\ell}(nS) \approx\ \frac{M_{\Upsilon}^2 \tan{}^2\beta}{8\pi
\alpha v^2}\biggl[1-\frac{M_{A^0}^2}{M_{\Upsilon}^2}\biggr] \approx
10^{-1}
\end{equation}
Let us remark that the $A^0-\eta_b$ mixing could make the
$A^0$ width larger than expected, yielding quite non-monochromatic
photons from radiative $\Upsilon$ decays. 
\end{itemize}

\subsubsection*{Theoretical arguments supporting the hypothesis of a
light CP-odd Higgs boson}

From a theoretical viewpoint, the existence of a light
pseudoscalar in the Higgs sector is expected in certain
extensions of the SM. As an especially appealing example, let us mention the 
next-to-minimal supersymmetric standard 
model (NMSSM)  where a gauge singlet is added to the
MSSM two-doublet Higgs sector, leading
to seven Higgs bosons, where five of them neutral of which
two are pseudoscalars \cite{gunion}. The associated
phenomenology has to be examined with
great attention in different experimental environments
\cite{Hiller:2004ii}. 

The NMSSM can show an approximate $R$-symmetry (in the limit
that the Higgs sector trilinear soft breaking terms
are small) or a U(1) Peccei-Quinn symmetry (in the limit that the cubic
singlet term in the superpotential vanishes). In either case, the lightest
CP-odd Higgs would be naturally light. This possibility can be extended to 
scenarios with more than one gauge singlet \cite{Han:2004yd}, 
and even to the MSSM with a CP-violating Higgs sector \cite{Carena:2002bb,
Dedes:2002er} as LEP bounds can be evaded \cite{Kraml:2006ga}. 

Moreover, even Little Higgs models have an extended structure 
of global symmetries
(among which there can appear $U(1)$ factors) broken
both spontaneously and explicitly, leading to possible light pseudoscalar
particles in the Higgs spectrum.
The mass of such pseudoaxions is in fact not predicted by
the model, and small values are allowed even of the
order of a few GeV \cite{Kraml:2006ga}. 

\subsubsection*{Some experimental results motivating the search
for a light Higgs boson in different BSM scenarios}

Either the SM or a non-standard Higgs boson has
been elusive in all the searches performed so far. However, 
different experimental measurements might already give some indications
about the existence of a light non-standard Higgs. Let us 
summarize several examples below:

\begin{itemize}

\item[a)] It is important to stress that a light CP-odd Higgs of mass below the
$b\bar{b}$ threshold has not been excluded by LEP 
in different scenarios (see ref.20 and references therein). 
Interestingly, this possibility
is consistent with the event excess detected in
the LEP data for a Higgs with SM-like $ZZH$ coupling in the
vicinity of 100 GeV, according to the NMSSM 
as emphasized in refs.21 and 22.
Furthermore, the choice of
these parameters under the LEP bounds yielding small
fine-tuning at small (large) $\tan{\beta}$ imply nearly always
(often) the existence of a relatively light SM-like Higgs boson
that decays into two light, perhaps very light, pseudoscalars.

\item[b)]  Light dark matter: It is possible that the neutralino is
extremely light (100 MeV to 10 GeV) and can annihilate at a
sufficiently rate through a light pseudoscalar boson so that the
correct dark matter relic density is obtained in the NMSSM 
\cite{Gunion:2005rw,Ferrer:2006hy}. Let us remark that the last
and unsuccessful search for light dark matter carried out by Belle
and CLEO \cite{Tajima:2006nc,Rubin:2006gc}
looking for upsilon invisible decays (following
the proposal made in ref.27 only restricts a vector 
mediator like an extra gauge boson (commonly dubbed U-boson), 
without affecting the possibility of a (pseudo)scalar mediator like 
a (CP-odd) Higgs boson. 

\item[c)] The $g$-$2$ muon anomaly would require a light CP-odd
Higgs to reconcile the experimental value with the theoretical result
when combining one-loop and two-loop contributions \cite{Cheung:2001hz}.
However, large theoretical uncertainties in the leading
order hadronic and hadronic light-light contributions as well from
experimental results make any conclusion controversial.

\item[d)] Let us note the fact that no $\eta_b$ state has been found so
far despite intensive and dedicated searches. Recently, CLEO carried out
a search for $\eta_b(1S)$ and $\eta_b(2S)$ states in hindered magnetic
dipole transitions from $\Upsilon(3S)$ with negative results. In
fact, one can speculate that this failure is due to a rather broad
pseudoscalar $^1S_0$ bottomonium state as a consequence of the new
physics contribution \cite{Sanchis-Lozano:2005di}. 
Notice also that other decay modes \cite{Rosner:2006jz},
used for seeking $\eta_b$, can have their BFs lowered with respect
to the SM expectations, thereby reducing the chances of $\eta_b$
to be observed through these decay channels. Besides, the 
$A^0-\eta_b$ mixing \cite{drees} could
alter the properties of both physical states, leading to
deviations from the SM expectations. Thus, even the mixed (i.e., physical)
$A^0$ state might be broader than expected due to its $\eta_b$ component,
with important consequences for its detection.
We will study this point in detail in a forthcoming paper.

\end{itemize}

\section{The Proposal}

The prospects to probe 
new physics by testing LU
have been discussed in several meetings of the quarkonium 
working group (QWG) and the main conclusions can be read in 
ref.31. In fact, this proposal is 
supported by the QWG: see http://www.qwg.to.infn.it/, where the action
items reproduced below can be found in the BSM section.

Indeed, we think it is 
extremely interesting that the current B factories could 
run at the $\Upsilon(3S)$ resonance
during a certain period of time  
in order to collect at least $\sim$10 $fb^{-1}$ so that LU 
can be tested with a statistical error below a few percent, as argued
below. Systematic uncertainties achieve the utmost importance
at this point since CLEO has reported a $5\%$ systematic error in the 
BF ratio \cite{Besson:2006gj}. As we shall see, the cascade
decay of $\Upsilon(3S)$ tagged by two charged pions could 
significantly improve this systematic
limit according to our estimates.  

In fact, Belle has already
performed an engineering run at $\Upsilon(3S)$ collecting
an integrated luminosity of $2.9\ fb^{-1}$, which could be used to test LU.
Moreover, the already existing data at $\Upsilon(4S)$
can be used for testing LU by exploiting ISR and dipion transitions as briefly
commented below.
We therefore propose to carry out the following program:
\begin{itemize}
\item[i)] Measurement of the chain decay where the dipion would tag
$\Upsilon(1S,2S)$, subsequently decaying into electrons or muons:
\[ \Upsilon(3S) \to \pi^+\pi^- \Upsilon(nS), \Upsilon(nS) \to
\ell^+\ell^-,\ \ \ n=1,2,\ \ \ell=e,\mu \] with the total branching
fraction as
\[ BF[\Upsilon(3S) \to \pi^+\pi^- \Upsilon(1S,2S)] \times
BF[\Upsilon(1S,2S) \to \ell^+\ell^-] \sim (4-8) \times 10^{-4} \] 
$\Upsilon(1S)$ has an advantage versus
$\Upsilon(2S)$ of not requiring the subtraction of cascade decays
from the latter into the former. However, $\Upsilon(2S)$ has
a larger BF and LU should be tested in its decays as well. 
On the other hand, muons should possibly be
preferable to electrons in the final state.

\item[ii)] Measurement of the chain decay where the dipion would tag
$\Upsilon(1S,2S)$, subsequently decaying into taus, as follows
\[ \Upsilon(3S) \to \pi^+\pi^- \Upsilon(nS), \Upsilon(nS) \to
\tau^+\tau^-,\ \ n=1,2 \]
The detection of taus could be done by looking at one-prong decays,
mainly focusing on its muonic decay mode. This decay mode should provide
important sources of both the statistical and systematic errors
for the BFs ratio to be tested.

\item[iii)] The LU in $\Upsilon(3S)$ leptonic decays can also be
tested by comparing the rates
\[ \Upsilon(3S) \to \ell^+\ell^-,\ \ \ \ell=e,\mu,\tau \]
Basically, this analysis should follow similar steps imposing
analogous selection criteria for accepted events as in CLEO 
\cite{Besson:2006gj}
although the accelerator and detector specifics should obviously
make them vary accordingly.
\end{itemize}

Although we have argued before that one can dispense with the
radiative photon in the process (2) (as we are primarily interested in 
testing LU ), let us stress, however, the relevance of detecting it to confirm 
or reject the existence of a Higgs particle mediating the decay. 
Therefore, selection cuts on events should be imposed taking into account 
this possibility, i.e., without a veto that could somehow spoil 
the detection of photons in the tauonic $\Upsilon$ decays.

On the other hand, the first and relatively easy
test of LU can be carried out based on 
data already collected at   
$\Upsilon(4S)$ at the B factories. We present several alternatives 
complementing each another as well as the dedicated run
at $\Upsilon(3S)$, as follows

\begin{itemize}

\item[iv)] The dipion decay of the 
$\Upsilon(4S)$ resonance can tag the decay, i.e.,
\[ \Upsilon(4S) \to \pi^+\pi^- \Upsilon(nS), \Upsilon(nS) \to
\ell^+\ell^-,\ \ n=1,2,\ \ \ell=e,\mu,\tau \] 
with the total BF of the order
\[ BF[\Upsilon(4S) \to \pi^+\pi^- \Upsilon(1S,2S)] \times
BF[\Upsilon(1S,2S) \to \ell^+\ell^-] \sim 10^{-6}-10^{-7} \]
for $\ell=e,\mu$  
which probably would require a super B factory 
\cite{Akeroyd:2004mj} for our goal of testing LU,
but might deserve further attention by currently
running experiments like Belle and BaBar.

\item[v)] A combination of ISR
and dipion transitions would allow 
to reach the $\Upsilon$ resonances as well from the
$\Upsilon(4S)$ energy according to the cascade decays:
\[ e^+e^-[\Upsilon(4S)]\ \to\ \gamma_{ISR}\Upsilon(3S,2S), \ 
\Upsilon(3S,2S)\ \to\ \pi^+\pi^-\Upsilon(2S,1S) \]
Hence, the leptonic decays 
\[ \Upsilon(2S,1S) \to  \ell^+\ell^-,\ \ell=e,\mu,\tau \]  
can be used again to compare the tauonic mode versus the electronic 
and muonic modes. The corresponding cross sections can be estimated to lie
in the range 15-60 $fb$ for the two abovementioned cases, leading to several 
thousands of events for the muonic decay and about one thousand events
for the tauonic decay, once the different efficiencies involved
in the process are taken into account.

\end{itemize}

\subsubsection*{Foreseen statistical error}

Let us note that the statistical error is likely to be dominated by the
tauonic decay mode.
Taking the CLEO analysis \cite{Besson:2006gj} as a reference,
where about $5 \times 10^6$ $\Upsilon(3S)$ decays were collected 
corresponding to $1.2$ $fb^{-1}$ on-resonance, 
and using the combined BFs of the
preceeding section (points i) and ii)) for the cascade decays 
$\Upsilon(3S) \to \pi^+\pi^- \Upsilon(1S,2S), \Upsilon(1S,2S) \to
\ell^+\ell^-$ ($\ell=e,\mu,\tau$), one
can naively infer that the statistical error of the ratio $R_{\tau/\mu}$
as a function of the integrated luminosity is approximately given
by
\[ \epsilon_{stat}\ \simeq\ \frac{0.07}{\sqrt{L}};\ \ \
\epsilon_{stat}\ \simeq\ \frac{0.07}{\sqrt{N_{days}}} \]
where $L$ stands for the integrated luminosity
in units $fb^{-1}$ and $N_{days}$ for the number of data taking days at the
$\Upsilon(3S)$ resonance, assuming that one day of
data taking corresponds to $\simeq$1$fb^{-1}$. For the direct
leptonic decays shown in iii), the statistical error almost should follow
the same rule. In particular, notice that by  
setting $L=2.9\ fb^{-1}$ (corresponding to the collected luminosity during 
the engineering run of Belle),
one obtains a foreseen statistical error $\simeq 4\%$.

\subsubsection*{Foreseen systematic error} Estimating the systematic error 
of $R_{\tau/\mu}$ in this proposal
is more difficult than the statistical one as it depends on
the experimental method, accelerator, and
detector characteristics, chiefly available to the Belle and BaBar 
Collaborations. Our aim is rather to obtain a
reasonable estimate in order to evaluate the feasibility
and significance 
of this proposal as compared to recent CLEO results reporting 
a systematic error $\epsilon_{syst} \simeq 5\%$ for the ratio
$R_{\tau/\mu}$ \cite{Besson:2006gj}. We calculate below the foreseen
systematic error according to the method based on the cascade decays:
\[ \Upsilon(3S) \to \pi^+\pi^- \Upsilon(nS), \Upsilon(nS) \to
\ell^+\ell^-;\ \ \ n=1,2,\ \ \ell=e,\mu, \tau \]
which should allow one to determine the ratio of 
leptonic BFs according to the following formula:
\[ \frac{BF[\Upsilon(nS) \to \tau
\tau]}{BF[\Upsilon(nS) \to
\ell \ell]}=\frac{N_{\tau\tau}\ \varepsilon_{\ell\ell}}
{N_{\ell\ell}\ \varepsilon_{\tau\tau}};\ \ \ n=1,2,\ \ \
\ell=e,\mu 
\]
where $N_{\ell\ell}$ represents the observed number of 
leptonic decays and 
$\varepsilon_{\ell\ell}$ represents the reconstruction efficiencies
with $\ell=e,\mu,\tau$. Let us remark that
the total number of events
needed for the evaluation of absolute leptonic BFs 
(like in ref.32)
cancel out in our BFs ratio, thereby
avoiding a source of systematic error. Notice also
that the decay chains for the tauonic mode and the
electronic and muonic modes share a large
common part (till the decay and detection of the final leptons) 
so one might expect an important cancellation of the
systematic uncertainty, e.g.,  
for the ratio in the Monte Carlo (MC) simulation
of the whole decay chain.

Indeed, the efficiencies $\varepsilon_{\ell\ell}$
can be obtained from a MC simulation
where the dipion transition has to be included using a certain model.
This is a similar situation to the study of the 
$\Upsilon(2S) \to  \Upsilon(1S)\pi^+\pi^-$ decay in ref.32,
where the Voloshin-Zakharov model \cite{Voloshin:1980zf}
was employed, leading to $\simeq$3$\%$
as the systematic uncertainty of the reconstruction efficiency. 
A similar estimate of the systematic error due to the MC
simulation was recently reported by Belle for the 
$\Upsilon(4S) \to \Upsilon(1S)\pi^+\pi^-$ decay \cite{unknown:2006sd}
using the Brown-Cahn model \cite{Brown:1975dz}.

Setting $1\%$ and $2\%$ as reference values
for the systematic errors of the 
trigger and tracking efficiencies, respectively,
and keeping conservatively a $\lesssim 3\%$ uncertainty
for the {\em ratio} of reconstruction efficiencies, 
the systematic error of  $R_{\tau/\mu}$ turns out to be 
\[ \epsilon_{syst}\ \lesssim \ 3.7\% \]
which represents an improvement with respect to the 
$5\%$ quoted by CLEO. 
Needless to say, a more accurate evaluation
remains to be made by the experimental teams
involved directly in the check of LU if this proposal were favorably
considered.

\section{Summary and final remarks}

In this proposal, we present a preliminary study about the
possibility of probing new physics by testing LU
to the few percent level in upsilon decays 
in a B factory running at $\Upsilon(3S)$. Furthermore, 
even with the machine running at $\Upsilon(4S)$,
initial state radiation in combination with dipion
transitions should permit reaching $\Upsilon(1S,2S,3S)$ resonances
to carry out the proposed test as well. 
An integrated luminosity of $\sim 10fb^{-1}$ at $\Upsilon(3S)$
should suffice to detect/exclude a light Higgs boson of mass about
10 GeV to the $95\%$ CL. The search for light dark matter 
can also be seen 
as a related issue to be carried out in a B factory
under similar run conditions \cite{McElrath:2005bp}.

From a theoretical viewpoint, there are arguments supporting the
existence of a light CP-odd particle in the Higgs spectrum of several
scenarios beyond the minimal extension of the SM. The parameter
regions allowed in these models are far more extensive than those
in the MSSM, 
with significant consequences for the phenomenology in colliders.
From an experimental viewpoint, there are already hints (LEP events
excess, g-2 anomaly,...) suggesting the
existence of a light pseudoscalar Higgs compatible with the LEP
bounds.

The observation of a Higgs boson of mass below the $b\bar{b}$
threshold, chiefly decaying into a tau pair if kinematically allowed, 
should be quite difficult in LHC experiments but relatively easy at a 
B factory. We want to stress the role to be played by Belle and BaBar 
in this regard. According to our estimates,
Belle could 
check LU using their 2.9 $fb^{-1}$ sample already collected in the
engineering run at $ \Upsilon(3S)$, 
with a foreseen statistical and systematic precision of the
order of the few percent each. Conceivably, BaBar can also perform a similar
analysis by using their $400\ fb^{-1}$ integrated luminosity
collected at $\Upsilon(4S)$ energy.
As a final remark, a super B factory
might test LU to an unprecented precision even when
running at 
$\Upsilon(4S)$ if systematics are well under control.

Lastly, let us stress the relevance of the (even negative) result
stemming from this test for constraining model parameters
in many scenarios beyond the SM,
regarding both the LHC and the prospects of the ILC and
a super B factory.

\section*{Acknowledgements}
I thank the organizers of the BNM2006 Workshop for their warm hospitality
and stimulating atmosphere by allowing so many interesting discussions. 
Special thanks to M. Hazumi for useful suggestions. I also
gratefully acknowledge J. Bernabeu, M. Cavalli-Sforza, 
J. Duboscq, J.J. Hernandez, N. Lopez,  
J. Papavassiliou, M. Pierini, and S. Tosi for valuable comments. 
Research under grant FPA2005-01678 and GVACOMP2006-214.

\end{document}